\documentclass[aps,superscriptaddress,twocolumn]{revtex4-2}
\usepackage{graphicx}
\usepackage{amsmath}
\usepackage{bm}
\usepackage{times}
\usepackage{lineno}
\usepackage{color}

\newcommand{\ems}{\sqrt{s_{NN}}}
\newcommand{\pt}{p_{\rm T}}
\newcommand{\Dr}{\Delta R}
\newcommand{\Dp}{\Delta P}

\begin{document}
\title{Light nuclei production in Au+Au collisions at $\sqrt{s_{NN}}$=3 GeV from coalescence model}

\author{Yue Xu}
\affiliation{Institute of Modern Physics, Chinese Academy of Sciences, Lanzhou, Gansu 730000}
\author{Xionghong He}
\affiliation{Institute of Modern Physics, Chinese Academy of Sciences, Lanzhou, Gansu 730000}
\author{Nu Xu}
\affiliation{Institute of Modern Physics, Chinese Academy of Sciences, Lanzhou, Gansu 730000}

\begin{abstract}
The nucleon coalescence model is one of the most popular theoretical models for light nuclei production in high-energy heavy-ion collisions. The production of light nuclei $d$, $t$, $^{3}$He, and $^{4}$He is studied using the transport model JAM with a simplified afterburner coalescence at $\ems=3$ GeV Au+Au collisions.
We scan the cut-off of phenomenological coalescence parameters, the relative spatial distance $\Delta R$ and momentum difference $\Delta P$, for formation of light nuclei by nucleon coalescence to reproduce the light nuclei $p_{\rm T}$ spectra measured by STAR experiment.
The results indicate a potential connection between the coalescence parameters and the binding energy as well as the diameter of these light nuclei.
\par\
\par\
\end{abstract}

\maketitle

%%%%%%%%%%%
\section{Introduction}
The nuclear matter under extreme pressure and temperature conditions is of particular interest as its property and evolution will shed light on understanding Quantum Chromodynamics (QCD). 
High-energy heavy-ion collisions have been used to create possibly deconfined quark matter, which could have existed a few microseconds after the Big Bang. Among the final states, the light nuclei production is a sensitive probe to their production mechanism and the properties of system evolution~\cite{PhysRevC.17.1051,CHEN20181}. 
They can be used to extract information on nucleon correlations and density fluctuations in heavy-ion collisions, which may provide crucial insights for the space-time evolution of the collision and searching for possible critical point~\cite{SUN2017103,CAINES2017121,Yu_2020,STAR:2022hbp}. 
Light nuclei may also abundantly appear in stellar objects such as supernova and binary neutron star mergers~\cite{PhysRevC.77.055804, PhysRevC.104.035801}. Their presence may impact the evolution and equation-of-state of these systems by affecting the transport coefficients in the dissipative process and the neutrino emission~\cite{PhysRevC.78.015806,Sedrakian2020}. 
Another reason to study light nuclei production in heavy-ion collisions is the investigation of anti-nucleus's origin in comic ray~\cite{PhysRevD.99.023016,PhysRevD.96.103021}. The AMS-02 experiment~\cite{TING201312} in the International Space Station may have observed anti-nuclei flux in space~\cite{anti-he3}. It is debated whether these events come from dark matter annihilation or anti-matter in space. The answer depends on the background estimates from $p-p$ and $p-A$ collisions~\cite{PhysRevD.99.023016,PhysRevD.96.103021}.

Currently, there are several popular but very different theoretical models describing the mechanism of light nuclei production in high-energy heavy-ion collisions. The statistical model depicts the light nuclei are thermally produced during the hadronization, and the total yields of light nuclei do not change after chemical freeze-out~\cite{PhysRevC.17.1051, PhysRevC.21.1301}. The statistical model has successfully described the yields of light nuclei and yield ratios of different light nuclei species~\cite{ANDRONIC2011203,ANDRONIC2021122176,PhysRevC.100.024911}. While the binding energy of light nuclei is around several MeV, the thermal model can not explain the survival of these loosely bounded nuclei in the fireball in which the temperature near chemical freeze-out is around 100 MeV~\cite{ANDRONIC2011203}.
The nucleon coalescence model assumes that light nuclei are formed at the late stage near the kinetic freeze-out of the fireball evolution via the coalescence of nucleons when these constituent nucleons are close to each other in both the coordinate space and momentum space~\cite{SATO1981153, PhysRevC.99.014901}. The coalescence picture has been used for understanding the light nuclei yields and flow from the STAR experiment at $\ems=3-200$ GeV~\cite{PhysRevC.99.064905,PhysRevC.94.034908,2022136941}.
Besides, dynamical formation and dissociation of light nuclei based on kinetic nuclear reactions have also been used to explore the light nuclei production for many years~\cite{DANIELEWICZ1991712,PhysRevC.80.064902}. Recently, with the inclusion of light nuclei size in the relativistic kinetic equations, the light nuclei yield in both $p+p$ and Au+Au (Pb+Pb) can be well described~\cite{2021arXiv210612742S}.

This paper discusses the light nuclei production in Au+Au collisions at $\ems=3$ GeV by utilizing a simple nucleon coalescence model. The nucleons are produced via the Jet AA Microscopic Transport Model(JAM)~\cite{PhysRevC.61.024901}. The interest of such calculations is to investigate the coalescence parameter dependencies on different light nuclei species. The light nuclei transverse momentum $p_{\rm T}$ spectra from the coalescence calculations are compared with the data measured by STAR  ~\cite{SciPostPhysProc.10.040}. This study will provide an improved understanding of coalescence calculations and light nuclei formation mechanisms in heavy-ion collisions.

%%%%%%%%%%%
\section{Light nuclei production}
The JAM model is designed to simulate relativistic nuclear collisions from the initial stage of nuclear collision to ﬁnal state interaction at finite and high baryon densities~\cite{PhysRevC.61.024901,10.1093}. In the model, the initial position of each nucleon is sampled by the distribution of nuclear density, and the nuclear collision is described by the sum of independent binary hadron-hadron collisions. At low energies ($\ems<4$ GeV), inelastic hadron-hadron collisions produce resonances that can decay into hadrons. All the established hadronic states and resonances can propagate in space and time and interact with each other by binary collisions. 
The JAM model has both cascade mode and mean-field mode. In the cascade mode, each hadron is propagated as in the vacuum between collisions with other hadrons. In the mean-field mode, the nuclear equation-of-state effects have been included through a momentum-dependent potential acting on the particle propagation~\cite{PhysRevC.94.034906}. The calculations from the mean-field mode have successfully described the light nucleus flow measurements at $\ems=3$ GeV Au+Au collisions, while the cascade mode failed to explain the data~\cite{2022136941}.
In our analysis, we use the JAM in its mean-field mode (incompressibility parameter $\kappa=$ 380 MeV) to generate the Au+Au collision events.

\begin{figure}[htbp]
  \centering
  \includegraphics[scale=0.42]{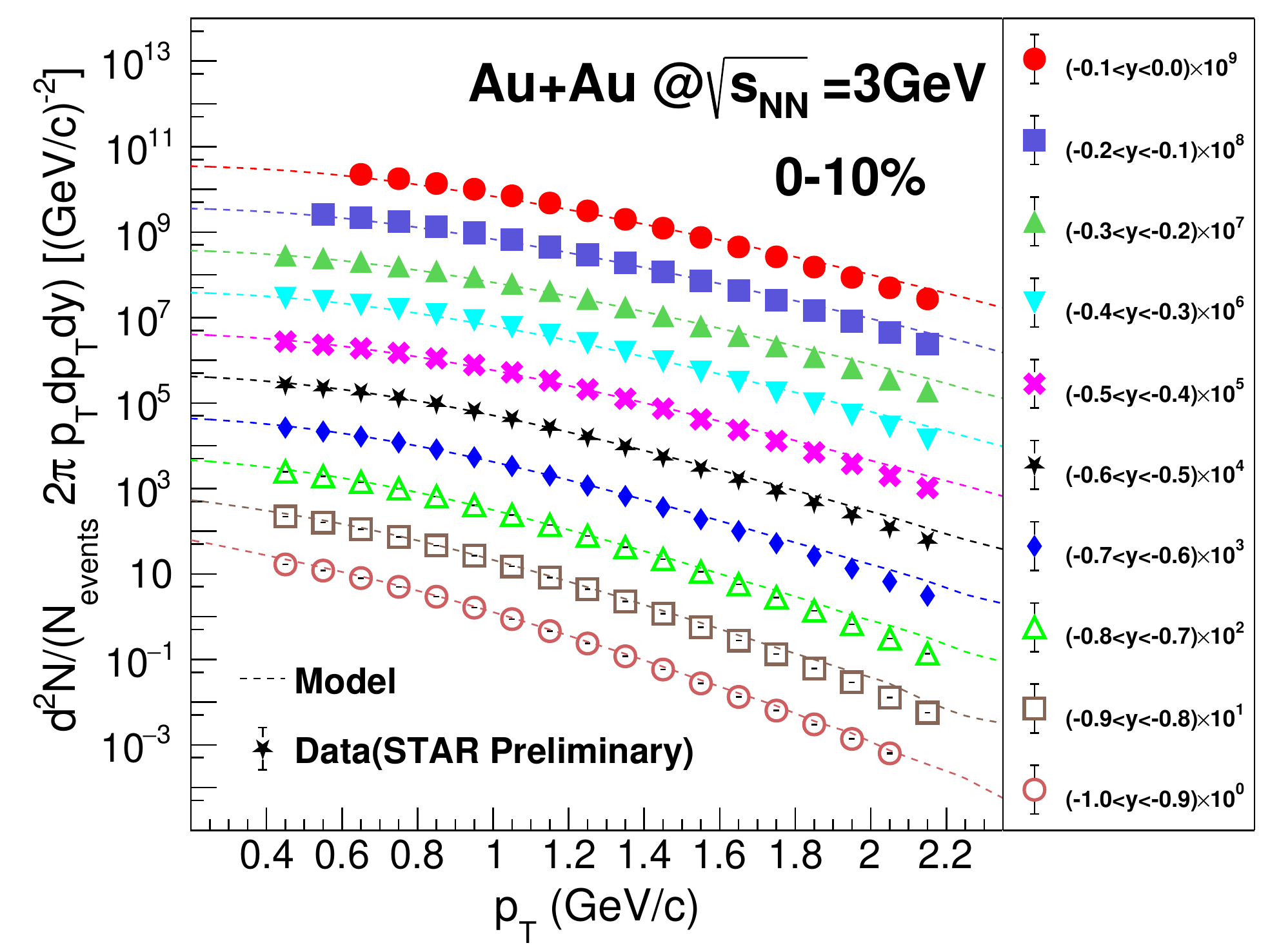}

  \caption{The proton $\pt$ spectra from the JAM calculations (solid lines) in various rapidity bins at 0-10\% Au+Au collisions at $\ems=3$ GeV. Markers represent data measured by STAR experiment.
  }
  \label{Fig.1}
\end{figure}

We obtain phase-space distributions for protons and neutrons from the JAM model for Au+Au collisions at $\ems=3$ GeV. Figure~\ref{Fig.1} shows the $\pt$ distribution of protons in various rapidity ($y$) intervals at 0-10\% most central collisions. The proton ($p$) $\pt$ spectra generally agree with the data measured by STAR experiment~\cite{SciPostPhysProc.10.040} at $\pt<1.4$ GeV/$c$ for all rapidity intervals. While the model calculations overestimate the data at higher $\pt$ region.

The JAM model does not produce light nuclei. Therefore, we utilize a simple afterburner coalescence to form deuteron ($d$), triton ($t$), $^3$He, and $^4$He. 
In an event from JAM, the positions and momentum of $p$ and neutron ($n$) are recorded at a fixed time of 50 fm/$c$ at which nearly all the nucleons are kinetically frozen out. Each $(p,n)$ pair is boosted to their rest frame, and then one obtains their relative space distance $\left|R_{1}-R_{2}\right|$ and relative momentum difference $\left|P_{1}-P_{2}\right|$. If both the values satisfy $\left|R_{1}-R_{2}\right|<\Dr$ and $\left|P_{1}-P_{2}\right|<\Dp$, where $\Dr$ and $\Dp$ are required values for $d$ formation, then this $(p, n)$ pair is marked as a $d$. For a $t$ formation, a $(p, n)$ pair is first formed according to $\Dr$ and $\Dp$ of $t$, and then one additional $n$ is included, and we calculate its relative space distance and momentum difference to the formed $(p, n)$ pair. The $^{3}$He and $^{4}$He are formed similarly with different $\Dr$ and $\Dp$.
This simple form of coalescence at a fixed time can be improved by using the wave function of light nuclei~\cite{PhysRevC.54.338,PhysRevC.102.044912}. However, the phase space coalescence worked successfully~\cite{2022136941} and gave similar results to that from wave function approach~\cite{PhysRevC.53.367}. Therefore, we utilize the simple coalescence method to provide some qualitative results for coalescence parameters $\Dr$ and $\Dp$ of different light nuclei species.

To determine the values of $\Dr$ and $\Dp$ that have the best descriptions for the light nuclei $\pt$ spectra, we carry out the scan of $\Dr$ and $\Dp$ for each light nuclei species, in which $\Dr$ is varied from 2 to 6 $fm$ with a step length of 0.2 $fm$ and $\Dp$ is varied from 0.1 to 0.5 Gev/$c$ with a step length of 0.04 Gev/$c$. 

Under the assumption that $n$ has the same phase space distribution as $p$ in the collisions, the invariant distribution of light nuclei can be expressed by the following equation
\begin{equation}
\begin{split}
E_{A} \frac{d^{3} N_{A}}{d^{3} p_{A}} & \propto\left(E_{p} \frac{d^{3} N_{p}}{d^{3} p_{p}}\right)^{Z}\left(E_{n} \frac{d^{3} N_{n}}{d^{3} p_{n}}\right)^{A-Z} \\
 & \approx \left(E_{p} \frac{d^{3} N_{p}}{d^{3} p_{p}}\right)^{A},
\end{split}
\label{eq1}
\end{equation}
where $A$ is the atomic mass number. Since the $\pt$ spectra of the proton from the JAM calculations overestimate the data at high $\pt$ as shown in Fig.~\ref{Fig.1}, the discrepancy with the model will be enhanced by a power factor of $A$ for light nuclei with a mass number of $A$ according to Eq.~(\ref{eq1}). To reduce the impact on the determination of coalescence parameters, the $\pt$ spectra for light nuclei are corrected following Eq.~(\ref{eq2}).
The yield ratio of data to the model calculation is obtained in a given $(\pt, y)$ cell of proton results, and then the light nuclei yield from the model calculation is corrected by $A$th power of the factor in the $(A\pt,y)$ cell.
\begin{equation}
\frac{d N_{\text {data }}^{A}}{d N_{\text {model }}}\left(A \pt, y\right)=\left(\frac{d N_{\text {data }}^{p}}{d N_{\text {model }}}\left(\pt, y\right)\right)^{A}.
\label{eq2}
\end{equation}
For light nuclei in a given rapidity interval, the $\pt$ range for applying the correction is determined by the $\pt$ coverage of the data points of proton.

%%%%%%%%%%%
\section{Results and discussion}
\begin{figure}[htbp]
  \centering
  \includegraphics[scale=0.45]{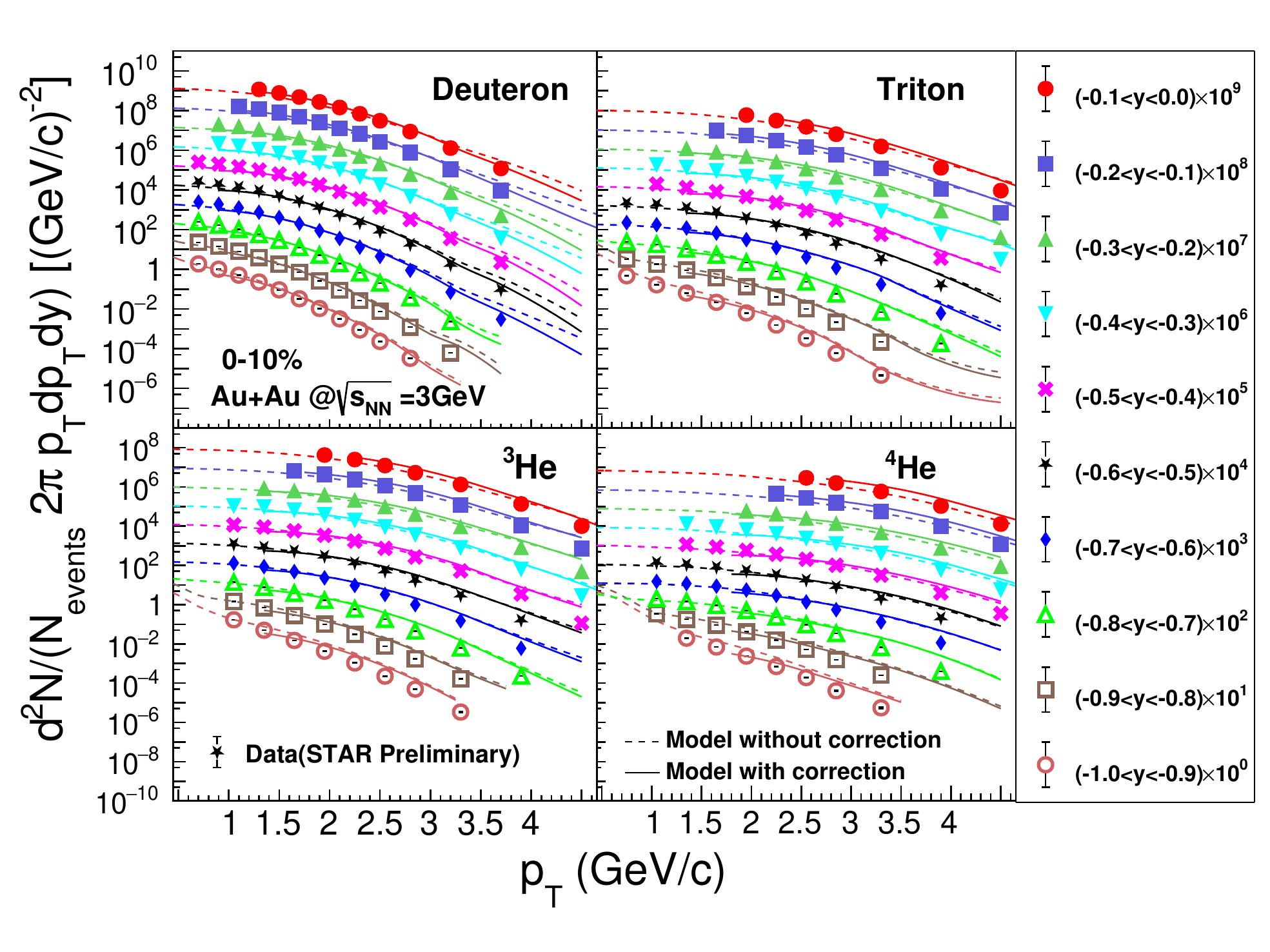}
  \caption{The $\pt$ spectra for $d$, $t$, $^3$He, and $^4$He data measured by STAR experiment and calculations from JAM model. Markers represent data. The dashed and solid lines represent the model calculations without and with corrections via Eq.~(\ref{eq2}), respectively. The cut-off for the correction result is caused by the low limit of proton $\pt$ in the data.}
  \label{Fig.2}
\end{figure}

Figure~\ref{Fig.2} shows the $\pt$ spectra for $d$, $t$, $^{3}He$, and $^{4}He$ in various rapidity bins in 0-10\% Au+Au collisions at $\ems=3$ GeV. 
The results of model calculations are obtained using the coalescence parameters $\Dr=4.0$ $fm$ and $\Dp=0.3$ GeV/c.
The distributions with and without the correction via Eq.~(\ref{eq2}) are shown by solid and dashed lines, respectively. 
After the correction, the model calculations can qualitatively reproduce the experimental data for all light nuclei species, including the high $\pt$ region. It supports the validity of using the simple coalescence approach for light nuclei production in heavy-ion collisions at several GeV.

Near the target rapidity $y=-1.045$, the fragmentation of the target ions may also contribute to the production of light nuclei. Currently, most of the transport models are unable to describe the production of the fragments in high-energy heavy-ion collisions. While it is found that our calculations can match the $\pt$ spectra of light nuclei at target rapidity with the same coalescence parameters as for the mid-rapidity at 0-10\% centrality, where the light nuclei are believed to be formed mainly through the nucleon coalescence. This agreement can be understood using a simple picture. In the collisions, the protons and neutrons near the target rapidity are less affected by the system evolution compared to those in the mid-rapidity; their momentum magnitude and direction will keep being close to each other. Thus, the nucleons near the target rapidity have a higher chance to combine, and the light nuclei production will be enhanced compared to the mid-rapidity, especially for $^4$He. While at peripheral collisions, the contribution of fragments near the target rapidity become more important~\cite{SciPostPhysProc.10.040} that the model calculations no longer match the data.

The light nuclei $\pt$ spectra in each rapidity interval obtained with a chosen ($\Dr, \Dp$) are compared to data using the $\chi^{2}$
\begin{equation}
\chi^{2}=\sum_{\pt} \left(\frac{v_{\rm data}-v_{\rm {model}}}{e_{\rm data}}\right)^{2},
\end{equation}
where $v_{\rm data}$ and $e_{\rm data}$ represent the light nuclei yield and its statistical uncertainty measured by STAR experiment, $v_{\rm {model}}$ is the model calculation for light nuclei yield with the correction by the Eq.(~\ref{eq2}). If the model has a better overall description of the data, the extracted $\chi^2$ will be smaller.

\begin{figure}[htp]
  \centering
  \includegraphics[scale=0.45]{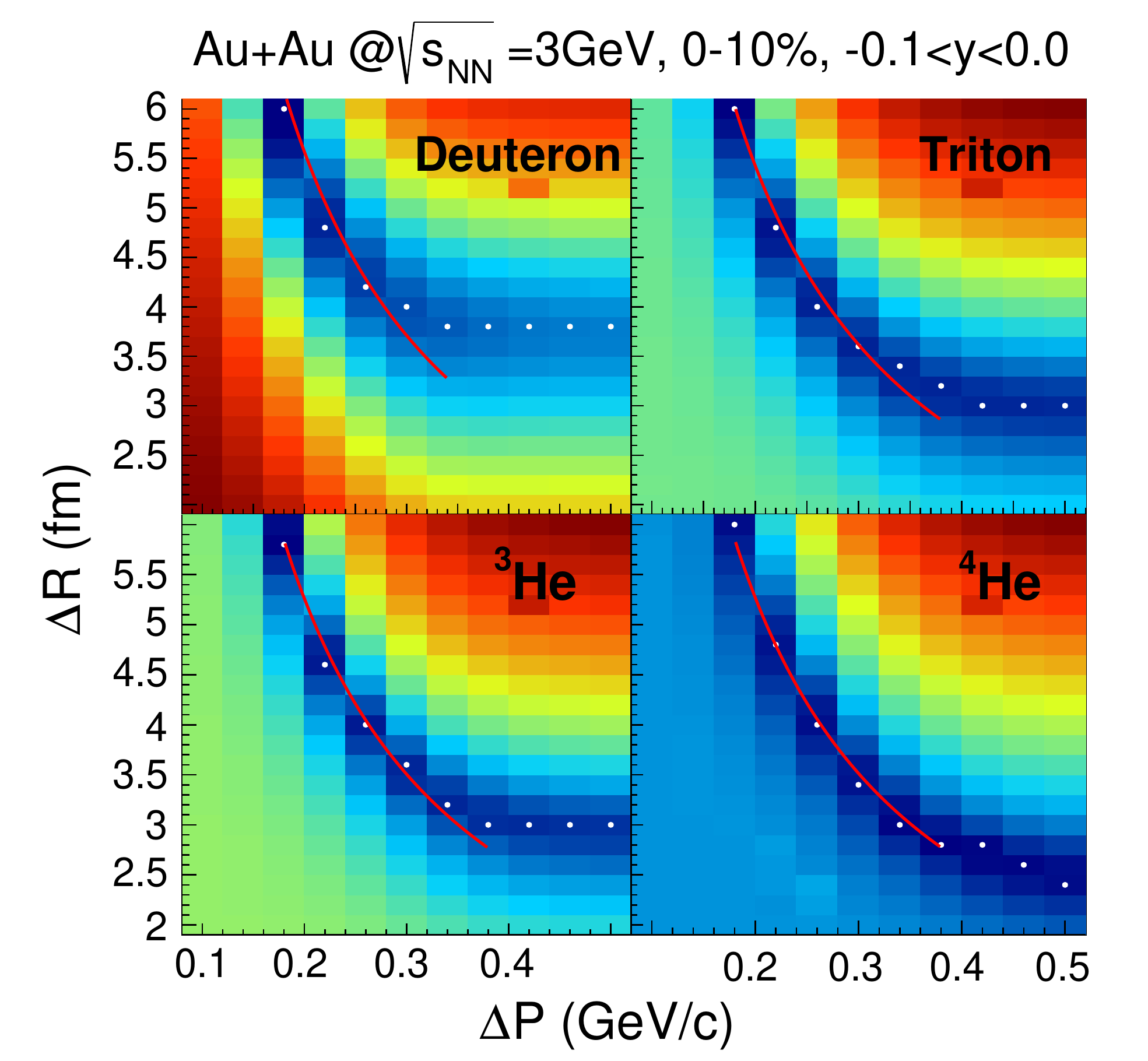}
  \caption{The dependencies of $\chi^2$ on coalescence parameters ($\Dr, \Dp$) for $d$, $t$, $^3$He, and $^4$He within $-0.1 < y <0$ in 0-10\% Au+Au collisions at $\ems=3$ GeV. The dots represent the minimum position of the $\chi^2$ for a given $\Dp$. The solid lines are the fits to these minimum values with an inversely proportional function.}
  \label{Fig.3}
\end{figure}

Figure~\ref{Fig.3} shows dependencies of the $\chi^{2}$ on the $\Dr$ and $\Dp$ at the mid-rapidity $-0.1<y<0$ in 0-10\% Au+Au collisions at $\ems=3$ GeV. The minimum position of the $\chi^2$ can be constrained by the $\Dr$ in a given $\Dp$ bin for all studied light nuclei species, especially for $d$ due to its large number of data points and small statistical errors. At low $\Dp$, the $\Dr$ for the minimum $\chi^2$ decreases with increasing $\Dp$, while the product $\Dr\cdot\Dp$ remains nearly unchanged, which the values are 1.2 for $d$ and $t$, and 1.15 for $^3$He and $^4$He. 
Following $\chi^2-$ minimization, the $\Dr$ is constant at high $\Dp$ region for $d$, $t$, and $^3$He, which means the light nuclei yields will not increase with increasing $\Dp$.
The value of ($\Dr, \Dp$) at the minimum $\chi^2$ are ($\Dr=5.6$ fm, $\Dp=0.22$ GeV/c) for $d$, ($\Dr=5.6$ fm, $\Dp=0.22$ GeV/c) for $t$, ($\Dr=5.2$ fm, $\Dp=0.22$ GeV/c) for $^3$He, and ($\Dr=3$ fm, $\Dp=0.42$ GeV/c) for $^4$He at $-0.1 <y<0$. For each light nuclei species, the $\chi^2$ distributions in other rapidity bins are very similar to the one in Fig.~\ref{Fig.3}, and the extracted ($\Dr, \Dp$) using $\chi^2-$ minimization has no strong dependence on particle rapidity. The average value of $\Dr$ and $\Dp$ in all rapidity bins are ($\Dr=5.78$ fm, $\Dp=0.219$ GeV/c) for $d$, ($\Dr=5.43$ fm, $\Dp=0.237$ GeV/c) for $t$, ($\Dr=5.24$ fm, $\Dp=0.242$ GeV/c) for $^3$He, and ($\Dr=4.585$ fm, $\Dp=0.3$ GeV/c) for $^4$He. A weak dependence of $\Dr$ and $\Dp$ on light nuclei rapidity implies that light nuclei are formed after the cascade stage of the reaction in high-energy heavy-ion collisions.

 \begin{figure*}[htbp]
  \centering
  \includegraphics[scale=0.85]{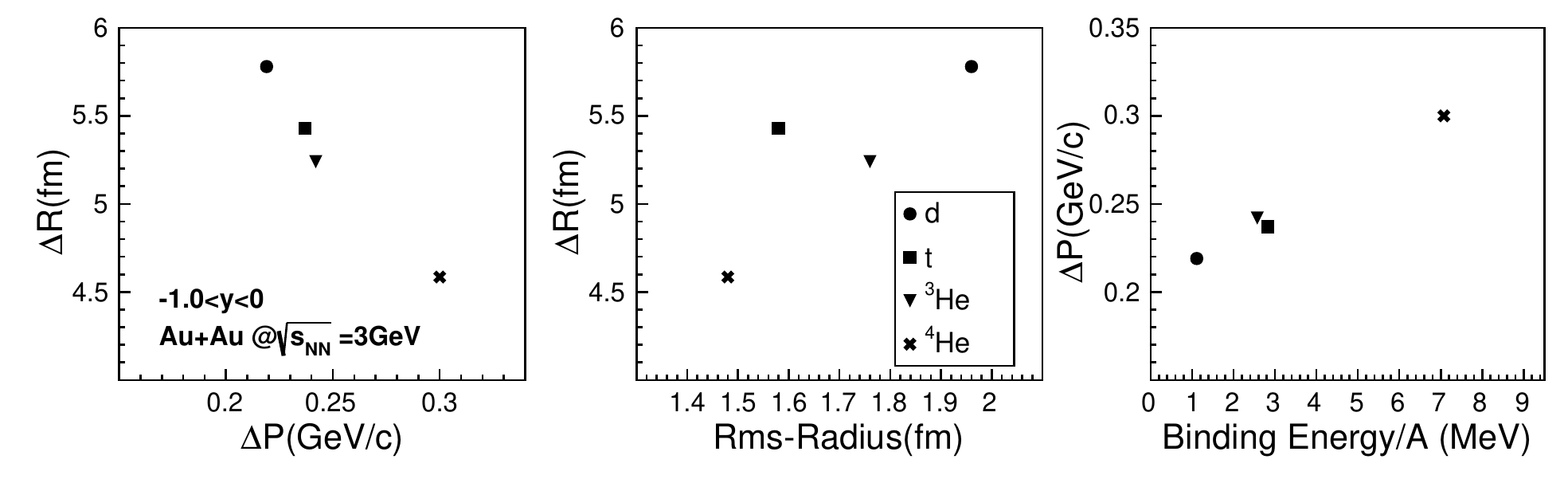}
  \caption{Left:$\Dr$ and $\Dp$ of the minimum $\chi^2$ in all rapidity bins for $d$, $t$, $^3$He, and $^4$He. Middle: The $\Dr$ as a function of nuclei rms diameter. Right: The $\Dp$ as a function of nuclei binding energy per nucleon.}
  \label{Fig.4}
\end{figure*}

Figure~\ref{Fig.4} shows the extracted $\Dr$ and $\Dp$ with the minimum $\chi^2$ describing the data in all rapidity bins at 0-10\% centrality, and their dependencies on nuclei diameter and binding energy~\cite{ROPKE201166}. Among the studied light nuclei species, the $d$ has the largest $\Dr$ and smallest $\Dp$, while $^4$He shows the opposite trend. In the middle and right panels of Fig.~\ref{Fig.4}, $\Dr$ is almost positively associated with nuclei diameter and $\Dp$ is positively correlated with nuclei binding energy. This suggests that nuclei with smaller binding energies can be formed with a higher upper limit for the relative momentum difference between its component nucleons. 
$^3$He and $t$ have very similar $\Dp$ and $\Dr$ as both their radius and binding energy are close. This investigation indicate that the radius and binding energy of light nuclei are crucial for their formation in heavy-ion collisions.

The $\Dr$ and $\Dp$ are supposed to be unique for a given nuclear species, thus it is expected to be independent of collision system and energy. It enables us to repeat the calculations for other collision energies with the same parameter sets and predict the light nuclei spectra and collective behavior. 
In this simple coalescence model, the formed light nuclei will sustain the collective flow of the produced nucleons, which are expected to be sensitive to the initial pressure gradient of the collision system. Light nuclei are heavier than nucleons, and their collective flow will have stronger energy dependence according to the coalescence model~\cite{2022136941} and thus is more sensitive to the change in pressure or the equation-of-state (EoS). Work in this direction is ongoing.

\section{Summary}
In summary, the light nuclei $d$, $t$, $^3$He, and $^4$He are formed via the phase space coalescence of nucleons produced by the JAM model in a mean field mode at $\ems=3$ GeV Au+Au collisions. We investigate the coalescence parameter cut-off, $\Dr$ and $\Dp$, by comparing the calculations with experimental data from the measurement by STAR experiment. For a given light nuclei species, a unique ($\Dr$, $\Dp$) is obtained, which can describe the $\pt$ spectra both at the mid-rapidity and at target rapidity at 0-10\% centrality. The result implies that at central collisions, the nucleons near the target rapidity may have higher coalescence probability as they are less affected by the system expansion compared to the mid-rapidity.
It is found that $\Dp$ and $\Dr$ are nearly positively correlated with nuclei binding energy and nuclei diameter, respectively. The result suggests that the radius and binding energy of light nuclei are crucial for their formation in heavy-ion collisions.

\bibliographystyle{apsrev4-2}
\bibliography{ref}  % Produces the bibliography via BibTeX.

\end{document}